\documentclass[aps,twocolumn]{revtex4}

\usepackage{epsfig}

\begin{document}

\title{Stochastic dynamics of dengue epidemics}

\author{David R. Souza$^1$, T\^ania Tom\'e$^1$, Suani T. R. Pinho$^2$,
Florisneide R. Barreto$^3$, and M\'ario J. de Oliveira$^1$}

\affiliation{$^1$Instituto de F\'{\i}sica, Universidade de S\~ao Paulo,
Caixa Postal 66318, 05314-970, S\~ao Paulo, Brazil \\
$^2$Instituto de F\'{\i}sica, Universidade Federal da Bahia, 
40210-340, Salvador, Brazil \\
$^3$Instituto de Sa\'ude Coletiva, Universidade Federal da Bahia,  
40110-140, Salvador, Brazil}

\begin{abstract}

We use a stochastic Markovian dynamics approach to describe 
the spreading of vector-transmitted diseases, like
dengue, and the threshold of the disease.
The coexistence space is composed by two structures
representing the human and mosquito populations.
The human population follows a susceptible-infected-recovered
(SIR) type dynamics and the mosquito population follows a 
susceptible-infected-susceptible (SIS) type dynamics.
The human infection is caused by infected mosquitoes and vice-versa
so that the SIS and SIR dynamics are interconnected.
We develop a truncation scheme 
to solve the evolution equations from which we get
the threshold of the disease and the reproductive ratio. 
The threshold of the disease is also obtained by performing numerical
simulations. We found that for certain values of the infection
rates the spreading of the disease is impossible whatever is 
the death rate of infected mosquito.




\end{abstract}

\maketitle

\section{Introduction}

Dengue is a vector-borne infectious disease with very complex
dynamics, whose spreading is a relevant problem of public health.
The disease is transmitted to human mainly by the mosquito
{\it Aedes Aegypti}. Many factors are determinant for the transmission
of dengue in urban centers such as the climatic conditions for the
vector proliferation, the human concentration and mobility.
Although many efforts are put in the development of a vaccine
against the four types of virus, until now the only available
strategy to reduce the spreading of the disease \cite{gubler97} 
is the control of the vector population. 
Therefore, it is very important to
analyze the effect of vector control avoiding the occurrence
of dengue epidemics. In this context, the inter-host modeling
of dengue dynamics and control may be a very useful tool for
helping its understanding and the establishment of vector control
strategies.

Different techniques and approaches are used to model the dynamics
of transmitted diseases 
\cite{keeling08,keeling99} such as deterministic differential 
equations \cite{brauer01}, stochastic dynamics \cite{souza10,tome10,tome11}, 
cellular automata \cite{rhodes96},
and complex networks \cite{pastor01}.
Concerning vector transmitted diseases including dengue modeling,
there are also different schemes and approaches 
\cite{keeling08,keeling99} based
on deterministic models of differential equations 
\cite{pinho10,fergunson99,massad01}, probabilistic
cellular automata \cite{martins00,ferreira06,santos09,botari11} and
complex networks \cite{silva07,ferreira07}. 
Although a description in terms of a
master equation defined on a lattice
has been used to investigate epidemic models of
direct transmitted diseases \cite{souza10}, 
this approach has not been explored in
the investigation of a vector-borne infectious disease.
This approach takes into
account in an explicit way the spatial structure of the environment
and, in contrast to mixing models, it predicts stochastic fluctuations and 
correlations in the number of individuals, features that are inherent
in real population dynamics. As we shall see, it allows a
definition of the basic reproduction ratio in terms of
conditional probability, which is non trivial only when correlations
are taken into account.

As in other works
\cite{pinho10,fergunson99,massad01,santos09},
the present approach is also motivated by actual data of dengue
epidemics \cite{barreto08}
in particular, by two outbreaks of dengue occurred in Salvador,
Bahia, Brazil in 1995 (without vector control) and 2002 (with
vector control). Those data had also motivated a previous 
analysis based on the basic reproductive ratio \cite{pinho10}.

\begin{figure}
\centering
\epsfig{file=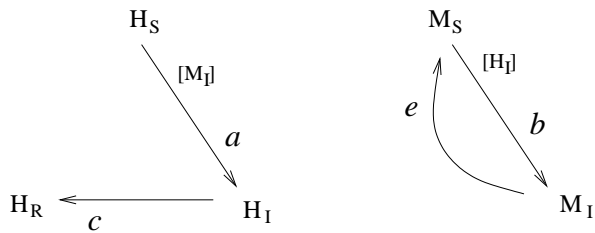,height=3cm}
\caption{Illustration of the reactions. 
Left panel: SIR structure. A susceptible human ($H_S$) becomes infected
($H_I$) through a catalytic reaction mediated by infected
mosquitoes ($M_I$). An infected human becomes recovered ($H_R$)
spontaneously and remains permanently in this state. 
Right panel: SIS structure. A susceptible mosquito ($M_S$) becomes infected
through a catalytic reaction mediated by infected humans.
An infected mosquito spontaneously becomes susceptible.}
\label{mod}
\end{figure}

The first dengue model was proposed by Newton and Reiter \cite{newton92}
in 1992 assuming a susceptible-exposed-infected-recovered (SEIR)
structure for humans and susceptible-exposed-infected (SEI) structure for 
mosquitoes due to the fact that the mosquitoes die before
being removed. This framework has been followed by other
continuous and discrete dengue models 
\cite{pinho10,fergunson99,massad01,ferreira06,santos09}.
Here we consider a simpler model, illustrated in figure \ref{mod},
assuming a susceptible-infected-recovered (SIR) structure for
humans and susceptible-infected-susceptible (SIS) structure for
mosquitoes. The infection of humans is due to mosquitoes
and the infection of mosquitoes is due to humans. In other
words the infection reactions, S$\to$I, on both structures are
catalytic and not autocatalytic as happens to the original
SIR and SIS models \cite{souza10,tome10,tome11}. 
The other reactions, I$\to$R in the SIR structure and
I$\to$S in the SIS structure, are spontaneous reactions. 
The mosquito structure has been simplified by suppressing the
death and the birth of mosquitoes which amounts to saying that
a dead mosquito is immediately replaced by a new-born 
(susceptible) mosquito. We are, therefore, assuming that
the number of mosquitoes remains constant throughout the
outbreak of the epidemics.

The features of our model that differ from that of Newton and Reiter
\cite{newton92} are as follows. Firstly, we do not distinguish between 
susceptible and exposed states both for humans and mosquitoes.
Secondly, deaths of humans are not considered since the human
life time is much larger than the period of the disease.
As to the deaths of the mosquitoes, they are implicit in 
the model in the following sense. The reaction $I\to S$
for the mosquitoes 
is to be interpreted as the death of an infected mosquito and 
the simultaneous birth of a susceptible mosquito. 
The major difference however rests on the use of a stochastic
lattice model, which takes into account the spatial 
distribution of humans and mosquitoes.

After setting up the master equation we develop two truncation schemes 
to solve the evolution equations from which we get the threshold of the
disease and the reproductive ratio. From the second truncation
scheme we found that, for a range of values of the infection rates,
the disease does not spread no matter how small is the rate at which
the infected mosquitoes disappear. This result is confirmed
by numerical simulations performed on a square lattice.

This paper is organized as follows. In section II, we introduce the model
and derive from the master equations the evolution equations for the
densities. We also define in this section the quantities that characterize
the spreading of the epidemics, the reproductive ratio and the size
of the epidemics. In section III, we develop the simplest truncation
scheme and show how some classic results are obtained. In section IV,
we introduce the second truncation scheme and set up the 
evolutions equations for densities and pair correlations. 
The stability analysis of these equations allows us to obtain
the threshold of epidemics from wich we get the phase diagram.
Section V is reserved to the numerical simulations of the model on
square lattice. Concluding remarks and discussion are placed in the
last section.

\section{Model and evolution equations}
\label{model}

The modeling of disease spreading that we consider here
corresponds to a continuous time stochastic Markovian process
defined on a lattice,
with periodic boundary conditions,
where the sites are occupied by human individuals or by mosquitoes. 
In order to properly describe the human and mosquito
populations we consider two sublattices of the whole lattice,
one for each population. 
The sublattices, named $H$ and $M$, are interpenetrating 
in such a way that the nearest neighbor sites of a site of one sublattice
belong to the other sublattice.
The number of nearest neighbors, the coordination number
$\gamma$, is the same for both sublattices.
Each site of the sublattice $M$ can be in 
one of two states, either occupied by a susceptible mosquito ($M_S$) 
or by an infected mosquito ($M_I$). Each site of the sublattice $H$ 
can be in one of three states, occupied by a susceptible human ($H_S$),
occupied by an infected human ($H_I$) or occupied by a recovered 
human ($H_R$). The system evolves in time according to the following
stochastic dynamics. 

Each site changes its state, independently of the others,
at waiting times distributed exponentially with rates that
depends on the state of the site and its neighborhood. 
(a) If a site is occupied by 
a susceptible human then it becomes infected with rate
$a$ times the fraction of infected mosquitoes in its neighborhood.
If the site is occupied by an infected human then it becomes
recovered spontaneously with rate $c$. Once recovered
the individual remains permanently in this state which means
that, if the site is occupied by a recovered human, it
remains unchanged. (b) If a site is occupied by
a susceptible mosquito then it becomes infected with a rate
$b$ times the fraction of infected humans in its neighborhood.
If the site is occupied by an infected mosquito it
spontaneously becomes susceptible with rate $e$.
Our simple model is therefore described by four parameters
$a$, $b$, $c$ and $e$. In the applications we will further
simplify by setting $a+e=b+c$.

For convenience we introduce a stochastic variable $\eta_i$ associated to
each site $i$ of the lattice that takes the values 0, 1, 2, 3, or 4
according to whether the site $i$ is occupied, respectively,
by a susceptible mosquito, an infected mosquito,
a susceptible human, an infected human or
a recovered human. The time evolution of the probability
distribution $P(\eta)$ of configuration $\eta=\{\eta_i\}$ is
governed by the master equation
\begin{equation}
\frac{d}{dt}P(\eta) 
= \sum_i \{ w_i({\cal A}^-_i\eta)P({\cal A}^-_i\eta) -w_i(\eta)P(\eta) \},
\label{master}
\end{equation}
where $w_i(\eta)$ is the transition rate from 
$\eta$ to $\eta'={\cal A}_i\eta$
and ${\cal A}_i$ is the operator that changes $\eta_i$ $\to$ $\eta_i'$
as follows: 0$\to$1, 1$\to$0, 2$\to$3, 3$\to$4; and ${\cal A}^-_i$
is the inverse of ${\cal A}_i$. The transition rate $w_i(\eta)$ 
is defined according to the rules stated above.
The time evolution of an average
$\langle f(\eta)\rangle=\sum_\eta f(\eta)P(\eta)$
is obtained from the master equation
and is given by
\begin{equation}
\frac{d}{dt}\langle f(\eta)\rangle 
= \sum_i \langle[f({\cal A}_i\eta) - f(\eta)]w_i(\eta)\rangle.
\label{average}
\end{equation}

Instead of using the full probability distribution $P(\eta)$ 
we consider an equivalent description in terms of the
various marginal probability distribution, obtained from $P(\eta)$.
The time evolution of the several marginal probability distributions
are obtained from the master equation (\ref{master})
and comprises a hierarchic 
set of  coupled equations which is equivalent to the master equation.
This approach is convenient because it allows to obtain
a solution of the set of equations by a truncation
scheme to be explained shortly. 
In what follows we assume
invariance of the properties by a translation of the lattice
by two lattice spacings so that a site of one sublattice goes into
another site of the same sublattice. Isotropy is also assumed. 
At $t=0$, a fraction $\epsilon$ of the mosquitoes sites,
which we consider to be very small, are infected and all the 
human sites are susceptible.

Let us denote by $P(\eta_i)$ the marginal
one-site probability that represents the probability that
a site $i$ is in state $\eta_i$ and by $P(\eta_i,\eta_j)$ the 
marginal two-site probability that represents the probability that site
$i$ is in state $\eta_i$ and a neighboring site $j$ is in state $\eta_j$. 
Other marginal probabilities are denoted in an analogous way.
The evolution equation of the one-site probability $P(1)$,
which represents the density of infected mosquitoes,
can be obtained from equation (\ref{average}) if we recall
that $P(1)=\langle \delta(\eta_i,1)\rangle$ and 
is given by
\begin{equation}
\frac{d}{dt}P(1) = b P(03) - e P(1).
\label{1}
\end{equation}
where we used the definition 
$P(03)=\langle\delta(\eta_i,0)\delta(\eta_j,3)\rangle$. 
The notation $\delta(x,y)$ stands for the Kronecker delta.
The time evolution equation for $P(0)$, the density of susceptible
mosquitoes, can be obtained from equation (\ref{1}) 
by using the property $P(0)+P(1)=1$.

The evolution equations for $P(2)$ and $P(3)$, the densities of
susceptible and infected humans, respectively, 
are obtained similarly from equation (\ref{average})
and are given by
\begin{equation}
\frac{d}{dt}P(2) = - a P(12),
\label{2}
\end{equation}
\begin{equation}
\frac{d}{dt}P(3) = a P(12) - c P(3).
\label{3}
\end{equation}
The evolution equation for $P(4)$, the density of recovered humans, is
\begin{equation}
\frac{d}{dt}P(4) = c P(3)
\label{4}
\end{equation}
and can also be obtained from equations (\ref{2}) and (\ref{3})
by taking into account the property $P(2)+P(3)+P(4)=1$.

To characterize the threshold of the epidemic it is convenient
to write equations (\ref{1}) for the density of infected mosquitoes
and (\ref{3}) for the density of infected individuals in the form
\begin{equation}
\frac{d}{dt}P(1) = bP(0|3) P(3) - e P(1),
\label{1b}
\end{equation}
\begin{equation}
\frac{d}{dt}P(3) = aP(2|1) P(1) - c P(3),
\label{3b}
\end{equation}
where $P(2|1)=P(12)/P(1)$ is the conditional probability of occurrence of
a susceptible individual given an infected neighboring mosquito
and $P(0|3)=P(03)/P(3)$ is the conditional probability of occurrence of
a susceptible mosquito given an infected neighboring individual.
Using the simplified notation $x=P(1)$ and $z=P(3)$
the set of equations (\ref{1b}) and (\ref{3b}) can be written as
\begin{equation}
\left(
\begin{array}{c}
dx/dt \\
dz/dt
\end{array}
\right) =
\left(
\begin{array}{cc}
-e & bP(0|3) \\
aP(2|1) & -c
\end{array}
\right)
\left(
\begin{array}{c}
x \\
z
\end{array}
\right).
\label{matrix}
\end{equation}
At the early stages of the epidemic
the cross transmission probabilities $P(0|3)$ and
$P(2|1)$ can be considered to be constant (independent of time)
and the set of equations
(\ref{1b}) and (\ref{3b}) becomes a linear set of equations.
A fundamental quantity that characterizes the spreading of the
disease is the so-called reproductive ratio $R_0$ which 
is defined here as follows
\begin{equation}
R_0 = \frac{ab}{ce} P(2|1)P(0|3).
\label{rep}
\end{equation}
The threshold of epidemic is determined by the largest eigenvalue
$\lambda$
of the matrix (\ref{matrix}), which is related to the reproductive ratio by
\begin{equation}
R_0 = \left(1+\frac{\lambda}{e}\right)\left(1+\frac{\lambda}{c}\right),
\label{eig}
\end{equation}
or by
\begin{equation}
\lambda = \frac12 \{ -(e+c) + \sqrt{(e-c)^2 + 4ec R_0} \}.
\end{equation}

According to the linear analysis the threshold of epidemic
occurs when the largest eigenvalue vanishes which happens,
according to equation ({\ref{eig}), when $R_0=1$. Moreover,
when $\lambda<0$ there is no transmission of disease.
According to equation (\ref{eig}) this happens when $R_0<1$.
The spreading of the disease occurs when $\lambda>0$,
that is, when $R_0>1$. The reproductive ratio characterizes
not only the threshold of the disease but also its strength.

An epidemic is usually characterized by the epidemic curve
defined as the number of cases occurring in unit time,
for instance, in a day or in a weak, plotted as a function
of time. In other terms it is the number of
susceptible individuals that are being infected per unit time.
In the place of number of individuals we may use the density
of individuals so that the epidemic curve is defined 
as the density of individuals that
are being infected per unit time $\zeta=-dy/dt$,
where $y=P(3)$ is the density of susceptible individuals. 
Using the initial conditions $x=\epsilon$, $y=1$ and $z=0$
where $\epsilon$ is a small quantity, $\zeta$
increases and then decreases in time and vanishes when
$t\to\infty$. 

The density of recovered individuals $\rho=P(4)$ increases
with time, as implied by equation (\ref{4}),
and approaches its maximum value in the limit
$t\to\infty$. This final density of
recovered individuals is the integral of the epidemic curve,
that is,
\begin{equation}
\int_0^\infty \zeta dt = \rho,
\end{equation}
obtained by integrating $\zeta=-dy/dt$, and by using
the initial condition $y(0)=1$ and the
result that the final density of infected individuals
vanishes, $z(\infty)=0$, and the constraint 
$\rho + y + z=1$. The final density of recovered individuals,
which is a measure of the size of the epidemics,
may be understood in this approach as the order parameter in the sense
that it vanishes in the non-spreading regime and is nonzero
in the regime where the disease spreads. Strictly speaking 
the vanishing only occurs when $\epsilon$ vanishes. Notice that,
the limit $\epsilon\to0$ should be taken after the limit $t\to\infty$,
which is the proper way to get the transition from the
spreading to non-spreading regime.

\section{Simple mean-field approximation}
\label{smf}

The three equations for the time evolution of $P(1)$, $P(2)$ and
$P(3)$ are not a set of closed equations because they depend on the
two-site probabilities $P(12)$ and $P(03)$. To get closed equations,
we firstly use a scheme, called simple mean-field (SMF)
approximation, which consists in writing a two-site probability
as the product of one-site probabilities,
that is, $P(\eta_i,\eta_j)=P(\eta_i)P(\eta_j)$.
Using the abbreviations $P(1)=x$, $P(2)=y$, and $P(3)=z$,
respectively, the densities of infected mosquitoes, 
susceptible individuals, and infected individuals,
this approximation gives $P(12)=xy$ and $P(03)=(1-x)z$ so that 
equations (\ref{1}), (\ref{2}) and (\ref{3}) are reduced
to the following forms
\begin{equation}
\frac{dx}{dt} = b (1-x) z - e x,
\label{81}
\end{equation}
\begin{equation}
\frac{dy}{dt} = -a x y,
\label{82}
\end{equation}
\begin{equation}
\frac{dz}{dt} = a x y - c z.
\label{83}
\end{equation}

To describe the spreading of the disease we consider
an initial condition such that all individuals are susceptible
so that $y=1$ and $z=0$. Moreover we consider a very small
density of infected mosquitoes, that is, $x=\epsilon$ where
$\epsilon<<1$. Then, we look for solutions for $x$, $y$ and
$z$ in the limit $t\to\infty$. In this limit the density
of infected mosquitoes and infected individuals vanish,
$x\to0$ and $z\to0$. 
If the density of recovered individuals
$\rho=1-y-z$ approaches a nonzero value than the disease
spreads. On the other hand if $\rho$ approaches zero
(when $\epsilon\to0$) the disease does not spread.
This means that the stationary solution $x=0$, $z=0$, $y=1$
($\rho=0$) is the solution corresponding to the non-spreading regime.

To obtain the stability of this solution
we linearize the above equations around this solution to
obtain
\begin{equation}
\frac{dx}{dt} = b z - e x,
\end{equation}
\begin{equation}
\frac{d\rho}{dt} = a x,
\end{equation}
\begin{equation}
\frac{dz}{dt} = a x - c z.
\end{equation}
A linear analysis of stability amounts to determine
the eigenvalues of the matrix composed by the coefficients
of the right hand side of these equations, given by
\begin{equation}
\left(
\begin{array}{ccc}
-e & 0  & b \\
 a & 0  & 0 \\
 a & 0  & -c 
\end{array}
\right).
\label{matrixsmf}
\end{equation}

One eigenvalue is zero and the others are the roots of
\begin{equation}
\lambda^2 + (e+c)\lambda + ec - ab = 0,
\end{equation}
that is,
\begin{equation}
\lambda_{\pm} = \frac12 \{ -(e+c) \pm \sqrt{(e-c)^2+4ab} \}.
\end{equation}
The solution is stable as long as $\lambda_+<0$, 
that is when $ec>ab$. The threshold of the spreading
occurs when 
\begin{equation}
\frac{ab}{ec} = 1.
\label{37}
\end{equation}
For $a+e=b+c$, the threshold line is described by 
$e=b$ as shown in the phase diagram of figure \ref{phase}.

The threshold obtained by the above linear analysis
can equivalently be obtained from the condition $R_0=1$,
that is, when the reproductive ratio $R_0$ given by equation
(\ref{rep}) equals one. In the present case of simple
mean-field approach $P(2|1)=P(2)$ and $P(0|3)=P(0)$.
The initial condition gives, $P(2)=1$ and $P(0)=1$ so that
\begin{equation}
R_0 = \frac{ab}{ce},
\end{equation}
which coincides with the condition (\ref{37}) when $R_0=1$.

\begin{figure}
\centering
\epsfig{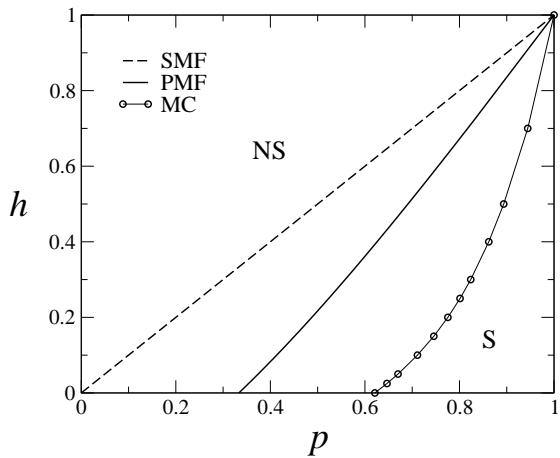}
\caption{Phase diagram in the plane $h=e/(e+a)$ versus $p=b/(b+c)$ 
for the case $e+a=b+c$, showing the
regions of spreading (S) and non-spreading (NS) of the disease for the
SMF approximation, PMF approximation for coordination number $\gamma=4$,
and numerical simulations (MC) on a square lattice.
When $e\to0$, the value of $b$ approaches zero for the SMF, the
value $p=1/3$ for the PMF and the value $p=0.621$
for numerical simulations.}
\label{phase}
\end{figure}

Near the threshold of epidemics the density of infected mosquitoes
is much smaller than unit and we may neglect $x$ in the first
term on the right hand side of equation (\ref{81}).
The evolution equation for $x$ becomes then
\begin{equation}
\frac{dx}{dt} = b z - e x,
\label{81a}
\end{equation}
which together with equations (\ref{82}) and (\ref{83})
allows us to determine explicitly the size of the epidemics $\rho$.
This is possible because these equations implies
a conservation law obtained as follows.
We start by defining the quantity $\phi=cx+bz$.
From equations (\ref{81a}) and (\ref{83}) its evolution
equation is given by 
\begin{equation}
\frac{d\phi}{dt} = -cex + abxy.
\label{84}
\end{equation}
The ratio of equations (\ref{84}) and (\ref{82}) gives
\begin{equation}
\frac{d\phi}{dy} = \frac{ce}{ay} - b,
\end{equation}
which can be integrated to give
\begin{equation}
cx + bz = \frac{ce}{a}\ln y + b (1-y), 
\end{equation}
which is the desired conservation law.
The constant of integration was obtained by remembering
that at $t=0$, $x=\epsilon\to0$, $y=1$ and $z=0$.

When $t\to\infty$, $x=0$, $z=0$ and $1-y=\rho$ so that
\begin{equation}
\ln (1-\rho) +  R_0 \rho = 0,
\end{equation}
where we used the relation $R_0=ab/ce$.
This equation determines the size of the epidemics
$\rho$ and may be written in the form
\begin{equation}
1-\rho = e^{-R_0 \rho}.
\end{equation}
This equation is the same equation obtained 
by Kendall \cite{kendall57} for the model introduced
by Kermack and McKendrick \cite{kermack27} to
describe the directed transmitted epidemics
\cite{keeling08,keeling99}
and also obtained by means of the simple mean-field approach to 
the SIR model on a lattice \cite{tome11}. Near the
threshold it is given by $\rho=2(R_0-1)$.

\section{Pair mean-field approximation}
\label{pmf}

Next we set up equations for the pair mean-field (PMF) approximation.
To this end we begin by writing the equations
for the two site probabilities.
The evolution equation for the probability $P(03)$ of
a site being occupied by a susceptible mosquito and a
neighboring site by an infected individual is given by
\[
\frac{d}{dt}P(03) = eP(13) - (c+rb)P(03) +
\]
\begin{equation}
+ (1-r)aP(021) - (1-r)bP(303),
\label{10}
\end{equation}
where $r=1/\gamma$, and $\gamma$ is the coordination number
of the lattice. The notation $P(\eta_i,\eta_j,\eta_k)$ stands
for the joint three-site probability where $i$ and $k$ are nearest
neighbor sites of $j$.  
The evolution equation for the probability $P(12)$ of
a site being occupied by an infected mosquito and a
neighboring site by a susceptible individual is given by
\[
\frac{d}{dt}P(12) = -(e+ra)P(12) 
\]
\begin{equation}
+ (1-r)bP(302) - (1-r)aP(121).
\label{15} 
\end{equation}
The equations for the other two-site probabilities,
$P(02)$, $P(04)$, $P(13)$, and $P(14)$ are not
necessary because they can be written in terms of $P(03)$,
$P(12)$, $P(1)$, $P(2)$ and $P(3)$.

Let us consider now the evolution equations for
$P(1)$, $P(2)$, $P(3)$, $P(03)$ and $P(12)$, given by
(\ref{1}), (\ref{2}), (\ref{3}), (\ref{10}) and (\ref{15}).
These five equations are not closed because
(\ref{10}) and (\ref{15}) include three-site probabilities. 
To get a set of closed equations
we now use a truncation at the level of two-site probabilities
\cite{satulovsky94,andersson00,kelly07}. This truncation 
amounts to use the following approximation for the three-site probability
\begin{equation}
P(\eta_i,\eta_j,\eta_k)=\frac{P(\eta_i,\eta_j)P(\eta_j,\eta_k)}{P(\eta_j)}.
\end{equation}
This approximation is used in equations (\ref{10}) and
(\ref{15}) to get a set of closed equations in the variables
$P(1)=x$, $P(2)=y$, $P(3)=z$, $P(03)=u$ and $P(12)=v$.
With this approximation the model is described by the set of five equations
\begin{equation}
\frac{dx}{dt} = b u - e x,
\label{21}
\end{equation}
\begin{equation}
\frac{dy}{dt} = - a v,
\label{22}
\end{equation}
\begin{equation}
\frac{dz}{dt} = a v - c z,
\label{23}
\end{equation}
\[
\frac{du}{dt} = e(z-u) - (c+rb)u +
\]
\begin{equation}
+ (1-r)a\frac{(y-v)v}{y} - (1-r)b\frac{u^2}{1-x},
\label{24}
\end{equation}
\[
\frac{dv}{dt} = -(e+ra)v +
\]
\begin{equation}
+ (1-r)b \frac{u (y-v)}{1-x} - (1-r)a\frac{v^2}{y}.
\label{25} 
\end{equation}

\begin{figure}
\centering
\epsfig{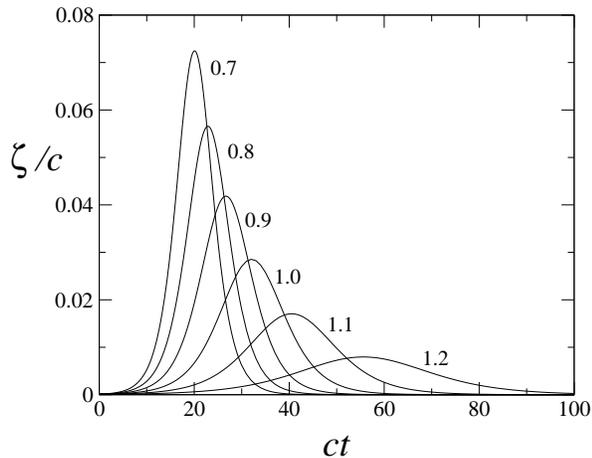}
\caption{Epidemic curves according to PMF approximation.
Each curve represents the density of individuals that are being
infected per unit time $\zeta=-dy/dt$ as a function of time
for the values of $e/c$ indicated and for $b/c=2$ and $a+e=b+c$.
The area below the epidemic curve $\zeta$ equals the final density
of recovered individuals $\rho$ and becomes neglible as one
approaches the threshold of epidemics.}
\label{temporal}
\end{figure}

We have solved the above set of equations using the
initial condition $x=10^{-4}$, $y=1$, $z=0$, $u=0$ and $v=10^{-4}$.
From the numerical solution we have obtained the
epidemic curve, that is the density of individuals that
are being infected per unit time, $\zeta=-dy/dt$.
Figure \ref{temporal} shows examples of the epidemic curve
obtained in the PMF approximation for $b/c=2$ and $a+e=b+c$.

To obtain the threshold of the spreading of the disease
we analyze the stability of the solution $x=0$, $y=1$,
$z=0$, $u=0$ and $v=0$, which characterizes the state
where the disease does not spread.
To get the stability of this equation we linearize
the evolution around this solution to obtain
\begin{equation}
\frac{dx}{dt} =  - e x + b u,
\label{31}
\end{equation}
\begin{equation}
\frac{d\rho}{dt} = a v,
\label{32}
\end{equation}
\begin{equation}
\frac{dz}{dt} =- c z + a v,
\label{33}
\end{equation}
\begin{equation}
\frac{du}{dt} = ez - (e+c+rb)u + (1-r)a v, 
\label{34}
\end{equation}
\begin{equation}
\frac{dv}{dt} =   (1-r)b u - (e+ra)v,
\label{35} 
\end{equation}
where $\rho=1-y-z$. A linear analysis of stability amounts to calculate
the eigenvalues of the matrix composed by the linear coefficients
of the right hand side of the above equations, given by
\begin{equation}
\left(
\begin{array}{rrrrr}
-e & 0  &  0 & b & 0\\
 0 & 0  &  0 & 0 & a\\
 0 & 0  & -c & 0 & a\\
 0 & 0  &  e & -(e+c+rb) & (1-r)a\\
 0 & 0  &  0 & (1-r)b & -(e+ra) 
\end{array}
\right).
\label{matrixpmf}
\end{equation}
Two eigenvalues are
$\lambda_1 = 0$ and $\lambda_2 = -e$. The others are the roots of
\[
-(\lambda+c)(\lambda+e+c+rb)(\lambda+e+ra)+
\]
\begin{equation}
+ ea(1-r)b + (\lambda+c)(1-r)^2ab = 0.
\label{roots}
\end{equation}
The line of stability is obtained by setting $\lambda=0$
in this equation, to get
\begin{equation}
-c(e+c+rb)(e+ra) + (1-r)abe + (1-r)^2abc = 0.
\label{43}
\end{equation}

As before we consider $b+c=a+e$ and write the equation that
describes the threshold line as
\begin{equation}
rc^2 + [e - (1-2r)b]c - (1-r)(b-e)e = 0.
\label{44}
\end{equation}
In figure \ref{phase} we show the line described by 
this equation for the case of coordination number $\gamma=4$.
This line represents the phase transition
between the spreading to non-spreading regime.
Below the transition line the non-spreading solution
becomes unstable giving rise to the spreading solution.
Comparing the pair and simple mean field approximation we
see that the inactive region of the phase diagram is larger
for the pair approximation, as can be seen in figure \ref{phase}. 

When $e=0$, the threshold of the disease occurs at
$b/c=r/(1-2r)$, a nonzero value. In terms of the quantity
$p=b/(b+c)$, used in the phase diagram of figure \ref{phase},
it occurs at $p=r/(1-r)=1/(\gamma-1)$. 
This result leads us  to conclude that there is a range of values
of $b/c$, as can be seen in figure \ref{phase},
for which there is no spreading of the disease 
whatever is $e/c$. This result is qualitatively distinct
from the SMF for which the threshold occurs at $b=0$ when $e=0$.
As we shall see this PMF prediction is confirmed by numerical
simulations. 

The threshold of epidemic obtained by the above linear analysis
can equivalently be obtained in terms of the reproductive ratio.
In the present case of pair mean-field approach 
$P(2|1)=P(12)/P(1)=v/x$ 
and $P(0|3)=P(03)/P(3)=u/z$ so that
\begin{equation}
R_0 = \frac{abuv}{ecxz}.
\label{rzpmf}
\end{equation}
The reproductive ratio $R_0$ depends on the parameters
$a$, $b$, $c$ and $e$ only through the ratios $e/c$, $b/c$ and $a/c$.
Indeed, if we substitute $\lambda$ given by equation (\ref{eig})
into equation (\ref{roots}) we get an equation
that gives $R_0$ in an implicit form. It is easy to see
that this equation contains the parameters
$a$, $b$, $c$ and $e$ only through the ratios $e/c$, $b/c$ and $a/c$.

It is worth mentioning that in the early stages of the
spreading of the epidemics the quantities $x$, $z$, $u$ and $v$ increase 
exponentially, that is, the increase in time of each of these
quantities is proportional to $e^{\lambda t}$,
where $\lambda$ is the largest eigenvalue of the matrix (\ref{matrixpmf}).
From this behavior
we see that the ratios $v/x=P(2|1)$ and $u/z=P(0|3)$
are independent of time at the early stage of the 
spreading of disease, and so is the reproductive ratio $R_0$.

An important aspect of our pair approach concerns the relation of the
epidemic curve with the conditional probabilities $P(2|1)$
and $P(0|3)$ which are treated exactly in this approach.
Remember that $P(2|1)$ is the conditional probability of the occurrence
of a susceptible individual in the presence of a infected
mosquito and $P(0|3)$ is the conditional probability of the occurrence of
a susceptible mosquito in the presence of a infected
individual. 

\section{Numerical simulations}

Numerical simulations were performed on a square lattice
according to the following rules.
At each time step, a site is chosen from a list of infected sites,
that is, a list of sites that are either occupied by
an infected mosquito or by an infected human individual. 
(i) If the chosen site is $M_I$ then with probability
$h$ it becomes $M_S$ and, with the complementary probability 
$1-h$, a neighboring site is chosen at random; if this
neighboring site is $H_S$ then it becomes $H_I$.
(ii) If the chosen site is $H_I$ then with probability
$q$ it becomes $H_R$ and with the complementary probability 
$p=1-q$, a neighboring site is chosen at random; if this
neighboring site is $M_S$ then it becomes $M_I$.
The time is then increased
by $1/N_I$ where $N_I$ is the number of sites in the list.
These rules are not the most general that one can
conceive from the original definition of the model
but are very simple and valid as long as $e+a=b+c$. 
Since the transition rates must be proportional to the transition
probabilities, it follows that
$a=\alpha(1-h)$, $b=\alpha(1-q)$, $c=\alpha q$, and $e=\alpha h$.
From these relations we see that $e+a=b+c=\alpha$ and may
write $h=e/(e+a)$, $q=c/(b+c)$ and $p=b/(b+c)$. 

At $t=0$ all sites of sublattice $H$ were
occupied by susceptible human individuals and
all sites of sublattice  $M$ were occupied by susceptible mosquitoes
except one site which is occupied by an infected mosquito.
We use lattice sizes sufficiently large 
so that the cluster of infected sites never reached the border of the lattice. 
The simulation was repeated a number of times, of the order of
a thousand, and the averages of relevant quantities were obtained.
For instance, we measured the mean number of infected human individuals
and the mean number of infected mosquitoes as functions of time.
The location of the critical point was obtained by assuming an algebraic
behavior of these quantities at the critical point. 

The results are shown in the phase diagram of
figure \ref{phase}. When $h=0$ ($e=0$), we have obtained
for $p$ a nonzero value, a result
qualitatively distinct from SMF and similar to PMF although
the value is a bit larger than that of PMF, namely, 
$p=0.621$. Therefore, our model predicts a range of values
of $p=b/(b+c)c$ for which the epidemics is impossible whatever is $h$. 

\section{Concluding remarks and discussion}

In this work we have applied stochastic dynamics to a dengue
bipartite lattice model to analyze the transition between epidemic
and non-epidemic states in terms of the probability of
human-mosquito and mosquito-human transmission and vector control
parameters. 
We have presented a precise definition of the reproductive rate
$R_0$ which is appropriate for systems described by a stochastic dynamics,
that characterizes the spreading of the disease, and
we have related it to the largest eigenvalue of the matrix associated
to the evolution equations. This definition can be generalized
to other types of disease transmission and seems to be promising
in the analysis of epidemics. According to our definition, the
reproductive rate is directed related to the conditional probability
of the occurrence of a susceptible human (mosquito) given
the presence in the neighborhood of an infected mosquito (human).
At the early stages of the epidemic, these conditional probabilities
are simply equal to the unity in the SMF approach
but are nontrivial in the PMF, what makes this approach a
richer description when compared to the SMF.
Another quantity that characterizes the epidemic is $\rho$,
the quantity that measures the
size of the epidemic and vanishes in the nonspreading regime.
It was also determined by means of the SMF and PMF approaches.

It is worth mentioning that the initial conditions we have used
in the mean-field approach is translational invariant.
If we have used an initial condition such that a finite number
of mosquito sites is infectious the initial state would not be
translational invariant and the mean-field calculation 
we have employed here would no longer be valid. For this initial condition
the disease may go to an early extinction even if $R_0>1$
and the growth of the epidemic may no longer be exponential.  

A qualitative relevant result that we have obtained
from the PMF approximation, and confirmed by numerical simulations,
is that for small values of $p$ there is no epidemics.
The minimum value for the spreading of the disease,
that occurs for $h=0$, is $p^*=1/(\gamma-1)$ for the PMF 
and $p^*=0.621$ for numerical simulations on a square lattice.
This result can be 
understood by relating the present model with percolation. 
It is well established that the SIR model has a close relation with
percolation \cite{tome10,tome11} so that it is to be expected
that the present model is also related to percolation growth. 

To appreciate the relation with percolation we consider
the spreading of the disease on a Cayley tree of coordination
$\gamma$ when $h=0$, starting from one single infected mosquito,
and observing the growing of the cluster of infected mosquitoes
and infected individuals. A site $M_I$ will remain forever in 
this state because $e=0$ and a site $H_I$ will eventually
become $H_R$ so that in the stationary state the percolating
cluster is formed by infected mosquitoes in sites belonging to
sublattice $M$ and recovered individuals in sites belonging to
sublattice $H$. In the process of cluster growing we may
ask for the probability that a site next to the border of the 
growing cluster will belong to the stationary cluster. If the site
belongs to the $H$ sublattice the probability is one. If the site
belongs to the $M$ sublattice, a calculation similar to that
of Tom\'e and Ziff \cite{tome10}, gives the value
\begin{equation}
p_M = \frac{p}{\gamma-(\gamma-1)p}.
\label{48}
\end{equation}
Therefore, we may say that the present model as defined on a Cayley tree
can be exactly mapped into an inhomogeneous site percolation on a Cayley
tree such that a site of sublattice $H$ is permanently active and
a site of sublattice $M$ is active with probability $p_M$.
The critical value of $p_M$ for percolation in such
a lattice is $1/(\gamma-1)^2$ instead of $1/(\gamma-1)$ 
as happens to homogeneous site percolation \cite{stauffer}.
If we substitute this value into (\ref{48}),
we get $p=1/(\gamma-1)$, which is the critical
value we have found by means of the PMF approximation. 

The relation to percolation allowed us to understand the existence 
of the minimum value of $p$ for the spreading of the disease
and therefore a range of values of infection rates for which the
epidemics is impossibe whatever is the death rate of infected 
mosquito. This result might be relevant in order to get the optimal
intervention scenario in the control of the disease.
In future work we will intend to analyze other important
issues such as the role of diffusion. Specifically, we wish to
know whether this scenario is preserved or not by the inclusion of diffusion. 

\section*{Acknowledgment}

STRP, TT and MJO thank CNPq and FAPESB (contract number PNX 006/2009)
for supporting the research. SRTP also thanks INCT (contract number
57386/2008-9). TT and MJO acknowledge INCT (contract number
573560/2008-0).


\end{document}